\documentclass[letterpaper,12pt]{article}
\usepackage[dvips]{graphicx}
\usepackage{subfigure}
\usepackage{amsmath}
\numberwithin{equation}{section}

\newlength{\dinwidth}
\newlength{\dinmargin}
\def\be{\begin{equation}}
\def\ee{\end{equation}}
\def\bea{\begin{eqnarray}}
\def\eea{\end{eqnarray}}

\begin{document}

\titlepage

\begin{flushright}
IPPP/05/58 \\
DCPT/05/116\\
September 2005 \\
\end{flushright}

\vspace*{4cm}

\begin{center}
%
{\Large \bf MHV techniques for QED processes}

\vspace*{1cm} \textsc{K.J. Ozeren and W.J. Stirling} \\

\vspace*{0.5cm} Institute for
Particle Physics Phenomenology, \\
University of Durham, DH1 3LE, UK \\
\end{center}

\vspace*{1cm}

\begin{abstract}
Significant progress has been made in the past year in developing
new `MHV' techniques for calculating multiparticle scattering
amplitudes in Yang-Mills gauge theories. Most of the work so far has
focussed on applications to Quantum Chromodynamics, both at tree and
one-loop level.
We show how such techniques can also be applied to abelian theories such
as QED, by
studying the simplest tree-level multiparticle process, $e^+e^- \to n
\gamma$.
We compare explicit results for up to $n=5$ photons using both the
Cachazo, Svrcek
and Witten `MHV rules' and the related Britto-Cachazo-Feng
`recursion relation' approaches with those using traditional spinor
techniques.

\end{abstract}

\newpage

\section{Introduction}

Until recently, the calculation of cross sections for the production of many particles (quarks, gluons, photons, etc.)
in high-energy collisions has been restricted by the technical difficulties associated with the evaluation
of the corresponding multiparticle Feynman diagrams. Significant progress has been made in the past year with the
development of new techniques, principally the `MHV rules' introduced by Cachazo, Svrcek and Witten (CSW) \cite{CSW} and the Britto-Cachazo-Feng (BCF) \cite{BCF} recursion relations, in which amplitudes are constructed from a new set of
building blocks -- Maximum Helicity Violating (MHV) amplitudes -- which themselves represent groups of Feynman diagrams
corresponding to particular external helicity configurations. In contrast to the usual approach, the amplitudes are expressed in terms of positive and negative chirality {\it spinors} ($\lambda$ and $\tilde\lambda$) and their corresponding products.
The MHV amplitudes have very simple expressions in terms of these spinor products, which replace the usual scalar products
$p_i \cdot p_j$ as the means by which the scattering amplitudes depend on the external particle four-momenta. Although first used to calculate tree amplitudes, these techniques have since been successfully applied at 1-loop \cite{loop}.

So far, the focus of attention has been on developing and exploiting techniques for QCD scattering processes, for example
$gg \to ng$, $gg \to q \bar q + ng$ etc., since such amplitudes are needed to estimate multi-jet cross sections at
hadron colliders. Indeed, the MHV rules \cite{CSW} were specifically developed for and applicable to massless Yang-Mills field theory.

In this paper we investigate whether similar techniques exist for abelian field theories, in particular QED. We choose
as our test-bed process the simplest QED tree-level multiparticle scattering process, $e^+e^- \to n\gamma$.\footnote{There is of course no tree-level QED analogue of the QCD process $gg \to ng$.}

In fact the $e^+e^- \to n\gamma$ process was already studied \cite{KS} almost twenty years ago at a time when `spinor techniques' for scattering amplitudes were being developed. The process was used to illustrate the power of these new techniques; extremely compact expressions were obtained for the production of arbitrary numbers of photons,
even taking the non-zero electron mass into account. We will refer to this as the KS approach. The expressions were specifically designed to allow for efficient
numerical computation of the amplitudes, and indeed the only limitation on the size of $n$ was due to the available
computing power at the time.

It is therefore interesting to see whether the more modern MHV-based techniques can improve on the calculational efficiency
of the original KS expressions. This can be measured, for example, by the compactness of the algebraic expressions and by the time taken to evaluate the spin-summed amplitude squared for one `event' corresponding to a random point in phase space.

The paper is organised as follows. In the next section we summarise the results of Ref.~\cite{KS} for the (massless)
$e^+e^- \to n\gamma$ scattering ampplitude.
 We then repeat the calculation using the QED-generalised version of the MHV technique, both in the
original MHV-rules and recursive (BCF) approaches.

\section{KS approach to $e^+e^-\to n \gamma$}

Consider the process
\be
e^-(p_a) + e^+(p_b) \to \gamma(k_1) +  \gamma(k_2) + ... + \gamma(k_n)
\ee
In the massless (electron) limit, helicity is conserved at each fermion-photon vertex and so the helicities
of the electron and positron will be opposite, $h_{e^-} = -h_{e^+}$. There are therefore $2^{n+1}$ distinct spin
amplitudes (two polarisations for each photon and $h_{e^-} = \pm$). In the traditional approach, the ($n!$) Feynman diagrams are obtained simply by joining the $n$ photons to the fermion line in all possible ways. Labelling the distinct polarisation states by $S = 1, ...,2^{n+1}$ gives an expression for the unpolarised cross section of
\be
d \sigma_n \; =\;  \frac{1}{F}\ \left[ d \Phi_n \right]\ \frac{1}{n!} \ \frac{1}{4} \sum_S \left| M_S \right|^2
\ee
where the terms on the right-hand side are the flux factor, the phase space volume element, the symmetry factor and the
spin-summed/averaged amplitude squared respectively.

The KS result is\footnote{This expression corresponds to the choice $h_{e^-}= - $. The corresponding  $h_{e^-}= + $ amplitudes are readily obtained using parity invariance.} \cite{KS}
\bea
M_S &=& e^n \left(\prod_{j=1}^{n} p \cdot k_j \right)^{-1/2} \sum_{D=1}^{n!}\;  \langle p_a \; a_1\rangle\; [p_b \; b_n] \nonumber \\
& &  \times  \prod_{i=1}^{n-1} \left\{  \frac{  \langle {\hat q}_i \; a_{i+1}\rangle\;  [{\hat q}_i\; b_i ] }{ q_i^2}
+  \frac{ \langle p \; a_{i+1}\rangle\; [p \; b_i]}{ 2p\cdot q_i}
\right\}
\label{eq:ks1}
\eea
where $p^\mu$ is an arbitrary light-like four vector and
\bea
a_i = p,\ b_i = k_i \ &\mbox{if}& \ h_i = + \; , \nonumber \\
a_i = k_i,\ b_i = p \ &\mbox{if}& \ h_i = - \; .
\eea
The sum in (\ref{eq:ks1}) is over the $n!$ distinct permutations $\hat k_1, \hat k_2, ...$ of the photon momenta
$k_1, k_2, ...$, from which internal four-momenta are defined by
\bea
q_i &=& \sum_{j=1}^{i} \hat k_j -p_a\; , \quad i=1, ..., n\quad (q_n \equiv p_b) \; ,\nonumber \\
\hat q_i &=&  q_i - \frac{q_i^2}{2p\cdot q_i}p_i\; , \quad (\hat q_i^2 = 0) \, .
\eea
The $\langle ij \rangle$ and $[ij]$ spinor products that appear in Eq.~(\ref{eq:ks1})
are defined in Appendix A.  The full expression for the amplitude for arbitrary $n$ can then be written in just a few lines of computer code. Note that the result for the amplitude squared is independent of $p^\mu$ (which is related to the choice of photon gauge) and this provides a powerful check on the calculational procedure.

\section{MHV and $\overline{\mbox{MHV}}$ Amplitudes}

Particular helicity amplitudes in Yang-Mills theory take on
unexpectedly simple forms. This is despite the considerable
computational effort involved in their calculation via the
standard Feynman diagram approach. Long expressions involving many
terms often simplify to a single term, or even vanish. At tree
level for example, purely gluonic colour-ordered scattering
amplitudes can be summarized as follows:\footnote{All particles are incoming, and the coupling constant factors have been omitted. The spinor products $\langle .. \rangle$ are defined in Appendix~A.}
\begin{eqnarray} \label{someamps}
\nonumber  A(1^+,2^+,...,n^+) &=& 0 \\
  A(1^+,2^+,...,i^-,...,n^+) &=& 0 \\
\nonumber  A(1^+,2^+,...,i^-,...,j^-,...,n^+) &=& \frac{\langle i \ j
\rangle^4}{\prod_{k=1}^{n} \langle k \ k+1 \rangle} \; .
\end{eqnarray}
So amplitudes with all the gluons having the same helicity vanish,
as do those with only one gluon having a different helicity to the
others. The third case above therefore corresponds to maximally helicity
violating (MHV) amplitudes. Their simple form was first conjectured by Parke and Taylor \cite{ParkeTaylor}, and later proven by Berends and Giele \cite{BerendsGiele} using a recursive technique. Note that
$A(1^+,2^+,...,i^-,...,j^-,...,n^+)$ above is called a `mostly plus' MHV amplitude, for
obvious reasons. Its `mostly minus' counterpart, which has two positive helicities and the remainder negative, is called an $\overline{\mbox{MHV}}$ amplitude, and can be obtained simply by interchanging $\langle .. \rangle \rightarrow [..]$.

\subsection{Photons}

The main difference when we consider QED is that there are no pure-photon tree-level amplitudes. There must always be (at least) one pair of fermions present, which must be of opposite helicity due to our convention
that all particles are incoming. Also, in contrast to non-Abelian theories there is no concept of colour ordering, so we will be concerned with full physical amplitudes rather than colour-ordered partial amplitudes. It is again the case that amplitudes with only one negative helicity particle (which must
be either the fermion or anti-fermion -- we shall take it to be the former) vanish,
\begin{equation}
A^{QED}(\overline{f}^+,f^-,1^+,2^+,...,I^+,...,n^+)=0\; .
\label{eq:QEDvanish}
\end{equation}
Here $i^+$ denotes a positive helicity photon with momentum $p_i$, and $f$, $\overline{f}$ denote fermion and anti-fermion respectively.
The MHV amplitudes take the following form (for massless fermions):
\begin{equation} \label{QEDMHV}
A^{QED}(\overline{f}^+,f^-,1^+,2^+...,I^-,...,n^+) =
\frac{2^{\frac{n}{2}} e^n \langle f \overline{f} \rangle^{n-2} \langle
f I \rangle^3 \langle \overline{f} I \rangle}{\prod_{k=1}^n
\langle f k \rangle \langle \overline{f} k \rangle}
\end{equation}
This is the fundamental MHV amplitude in QED, and as before it consists of only a single term. The factor $e^n$ is the gauge coupling constant, which we will normally omit in what follows. It is possible \cite{Mangano_qed} to obtain the amplitude in (\ref{QEDMHV}) by symmetrizing colour-ordered non-Abelian amplitudes,
\begin{equation}
A^{QED}(\overline{f},f,1,2,3,...,n) = A(\overline{f},f,1,2,3,...,n) + A(\overline{f},f,2,1,3,...,n) + ... \; .
\end{equation}
Each term on the right-hand side is a colour-ordered MHV (Parke-Taylor) QCD amplitude, and we sum over $n!$ permutations of $n$ gluons.  A factor of $2^{\frac{n}{2}}$ must also be included to take account of different generator normalizations -- our QED generators are normalized to $1$.
It should be noted that in writing (\ref{QEDMHV}) in this particular way we have made an apparently arbitrary choice of phase. Since the phase of a full (i.e. not partial) amplitude is not a physical observable, any of the $\langle..\rangle$ products in (\ref{QEDMHV}) could, naively, be replaced with the corresponding $[..]$ product. We will come back to this point later. It is worth mentioning that due to parity invariance the amplitude with all the helicities flipped has the same magnitude as (\ref{QEDMHV}) above. Also, one can use charge conjugation invariance to switch the fermion and anti-fermion.

We can write (\ref{QEDMHV}) in a physically more illuminating way, emphasizing the pole structure:
\begin{eqnarray}
A(\overline{f}^+,f^-,1^+,2^+...,I^-,...,n^+) &=& \frac{\langle f I
\rangle^3 \langle \overline{f} I \rangle}{\langle f \overline{f}
\rangle^2} \prod_{k=1}^n \frac{e\sqrt{2} \langle f \overline{f}
\rangle}{\langle f k \rangle \langle \overline{f} k \rangle} \\
    &=& \frac{\langle f I
\rangle^3 \langle \overline{f} I \rangle}{\langle f \overline{f}
\rangle^2}  \prod_{k=1}^n S_k\; .
\end{eqnarray}
It is a fundamental result of general quantum field theories that
scattering amplitudes have a universal behaviour in the soft (gauge boson)
limit. When all components of a particular photon's momentum are taken to zero, the amplitude factorizes into the amplitude in the absence of that photon multiplied by an `eikonal factor',
\begin{equation}
S_k = \frac{e\sqrt{2} \langle f \overline{f}
\rangle}{\langle f k \rangle \langle \overline{f} k \rangle}\; .
\end{equation}
The form of this factor is universal. Since the QED MHV amplitude is just a single term,
it follows that the eikonal factors must be present as factors -- and indeed they are.

\section{The MHV Rules}
There has been much recent progress in calculating scattering amplitudes in perturbative Yang-Mills theory. Cazacho, Svrcek and Witten \cite{CSW} introduced a novel diagrammatic technique, known as the `MHV rules', in which maximally helicity violating
(MHV) amplitudes are used as vertices in a scalar perturbation
theory. These vertices are connected by scalar propagators
$1/p^2$. This arrangement vastly reduces the number of diagrams
that must be evaluated relative to the traditional Feynman rules case.

Although the original CSW paper dealt only with purely gluonic amplitudes, the formalism has been successfully extended to include quarks \cite{Khoze,Glover}, Higgs \cite{Higgs} and massive gauge bosons \cite{Gauge Bosons}. In this paper we will use (\ref{QEDMHV}) to apply the MHV rules to QED processes, and derive relatively simple formulae for four and five photon amplitudes (an electron and positron are understood to be present also).

In order to use MHV amplitudes as vertices, it is necessary to
continue them off-shell, since internal momenta will not be
light-like. We need to define spinors $\lambda$ for the internal lines. The convention established in \cite{CSW}, which we shall follow, defines $\lambda$ to be
\begin{equation} \label{offshell}
\lambda_a = p_{a\dot{a}}\eta^{\dot{a}}
\end{equation}
for an internal line of momentum $p_{a\dot{a}}$, where $\eta^{\dot{a}}$ is arbitrary. The same $\eta$ must be used for all internal lines and
in all diagrams contributing to a particular amplitude. In
practice, it proves convenient to choose $\eta$ to be one of the conjugate (opposite chirality) spinors
$\widetilde{\lambda}$ of the external fermion legs. Note that for external lines, which remain on-shell, $\lambda$ is defined in the usual way (see Appendix A).

Having defined the MHV amplitudes, and the manner in which they
are to be continued off-shell, we are now in a position to
calculate non-MHV amplitudes. These are simply those with more
than two negative-helicity particles.
\subsection{Simple Examples}
As a first example let us calculate $A(\overline {f
^+_1},f^-_2,3^-,4^-)$.\footnote{Note the change in notation -- the spinor representing the fermion is now denoted $2$ (not $f$) and the spinor representing the anti-fermion is now denoted $1$ (not $\overline f$). Also, for clarity we will now omit the coupling constants.}  This is expected to vanish, see (\ref{eq:QEDvanish}). There are two
contributing MHV diagrams, though they differ only by a permutation of
photons. Note that the external legs are not constrained to be positioned cyclically as in the case of colour-ordered partial amplitudes. The absence of a pure-photon vertex means that the internal lines of MHV diagrams for QED processes with two fermions can only be fermionic. The contribution of the diagram in Figure 1 can be written down immediately as
\begin{figure}
\centering
\includegraphics{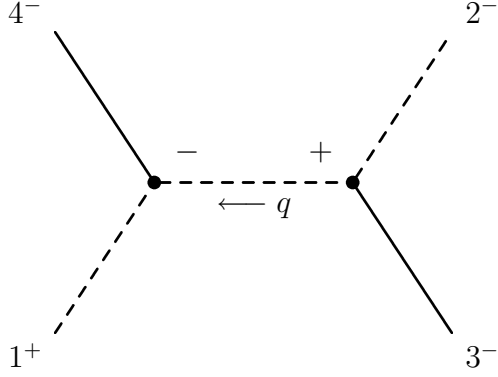}
\caption{Diagram contributing to $A(\overline
f{^+_1},f^-_2,3^-,4^-)$. Fermion lines are dashed, photon lines are solid. All particles are incoming. There is also a
similar diagram with photons 3 and 4 interchanged.}
\end{figure}
\begin{equation}
\sqrt{2} \frac{\langle \lambda_q \ 4 \rangle ^2}{\langle \lambda_q \ 1 \rangle} \frac{1}{q^2}  \sqrt{2} \frac{\langle 2 \ 3 \rangle ^2}{\langle 2 \ \lambda_{q} \rangle} \; ,
\end{equation}
where $\lambda_q$ is the spinor representing the internal line of momentum $q$. This expression is simply a product of two MHV vertices and a propagator. Using (\ref{offshell}) we can evaluate the spinor products involving $\lambda _q$:
\begin{eqnarray}
\langle \lambda_q \ 4 \rangle &=& \langle 4 \ 1 \rangle \ \phi_1\; , \nonumber \\
\langle \lambda_q \ 1 \rangle &=& \langle 1 \ 4 \rangle \ \phi_4\; , \nonumber  \\
\langle 2 \ \lambda_{q} \rangle &=& \langle 2 \ 3 \rangle \ \phi_3\; ,   \nonumber  \\
q^2 &=& (k_2 + k_3)^2   \; ,   \nonumber \\
&=& \langle 2 \ 3 \rangle [2\ 3] \; .
\end{eqnarray}
Here $\phi_i = [\eta \ i]$ is a function of the (arbitrary) spinor $\eta$. Simplifying, we find
\begin{equation}
-2 \frac{\langle 4 \ 1 \rangle}{[2 \ 3]} \frac{\phi_1^2}{\phi_3 \phi_4} \; .
\end{equation}
To this we must add the contribution from the diagram with photons 3 and 4 interchanged, namely
\begin{equation}
-2 \frac{\langle 3 \ 1 \rangle}{[2 \ 4]} \frac{\phi_1^2}{\phi_4 \phi_3}\; .
\end{equation}
When we add these two terms we find that momentum conservation, which
can be expressed as $\langle 1 \ 4 \rangle [ 4 \ 2 ] + \langle 1 \
3 \rangle [ 3 \ 2 ] =0 $, ensures that the sum vanishes, so that
\begin{equation}
A(\overline {f ^+_1},f^-_2,3^-,4^-)=0
\end{equation}
as expected.

The next test is to work out the $\overline{\mbox{MHV}}$ amplitude
$A(\overline {f ^+_1},f^-_2,3^-,4^-,5^+)$. There are four
diagrams (see Figure 2), though once again our task is simplified because there are
only two independent expressions to work out, the rest being obtained by appropriate permutations.
We find that
\begin{eqnarray}
  M &=& 2\frac{\langle \lambda_q \ 4 \rangle ^2}{\langle \lambda _q \ 5
\rangle \langle 1 \ 5 \rangle} \frac{1}{q^2} \sqrt{2}\frac{\langle 2 \ 3
\rangle ^2}{\langle 2 \ \lambda_{q} \rangle} \nonumber \\
   &=& 2^{\frac{3}{2}}\frac{[\langle 4 \ 1 \rangle \phi_1 + \langle 4 \ 5 \rangle \phi_5]^2}{[\langle 5 \ 1 \rangle
   \phi_1 + \langle 5 \ 4 \rangle \phi_4] \langle 1 \ 5 \rangle [2 \ 3] \phi_3}  \\
\end{eqnarray}
and
\begin{eqnarray}
  N &=&   \sqrt{2}\frac{\langle \lambda_r \ 4
\rangle ^2}{\langle \lambda_r \ 1 \rangle} \frac{1}{r^2} 2\frac{\langle 2 \ 3 \rangle ^2}{\langle \lambda _{r} \ 5
\rangle \langle 2 \ 5 \rangle}     \nonumber \\
   &=& -2^{3/2}\frac{\langle 2 \ 3 \rangle^2 \phi_1^2}{[\langle 5 \ 2 \rangle
   \phi_2 + \langle 5 \ 3 \rangle \phi_3] \langle 2 \ 5 \rangle [1 \ 4] \phi_4}  \; .
\end{eqnarray}
\begin{figure}
\centering
    \subfigure[M]{
    \label{fig:subfig:a}
    \includegraphics[width=2.5in]{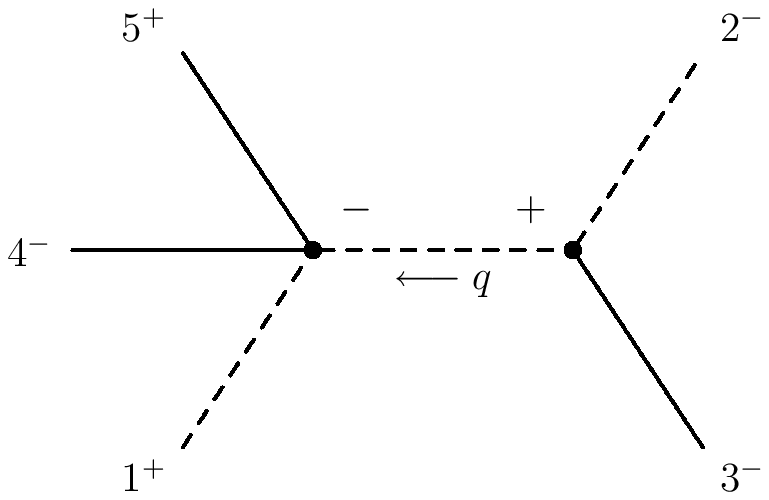}}
    \subfigure[N]{
    \label{fig:subfig:bb}
    \includegraphics[width=2.3in]{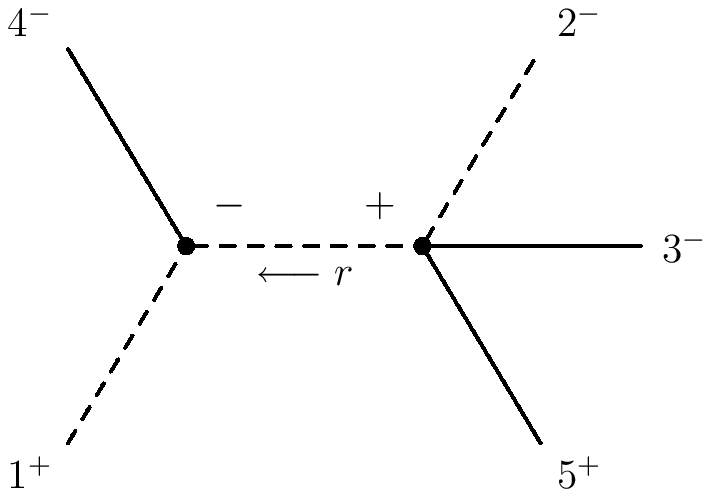}}
\caption{Diagrams contributing to $A(\overline {f
^+_1},f^-_2,3^-,4^-,5^+)$. The $3 \leftrightarrow 4$ permutation of each also contributes.}
\end{figure}
If we choose $\eta = \tilde{\lambda_1}$ then $\phi_1 = 0$ and so $N$ and its $3 \leftrightarrow 4$ permutation vanish. We are left with two terms which, after again invoking momentum conservation, simplify to
\begin{equation} \label{mhvbarresult}
A(\overline {f ^+_1},f^-_2,3^-,4^-,5^+) = -2^{\frac{3}{2}}\frac{[1 \ 2] [1 \ 5]^3 [2 \ 5]}{\prod_{k=3}^5 [1 \ k] [2 \ k]} \; .
\end{equation}
Inspection of the corresponding MHV amplitude (\ref{QEDMHV}) shows that this result has the correct magnitude. Having made a particular choice of phase for the MHV amplitudes in (\ref{QEDMHV}), a definite phase emerges for their $\overline{\mbox{MHV}}$ counterparts. The former were chosen to be holomorphic functions of the $\lambda$'s of the external legs -- they contain only $\langle .. \rangle$ products. The latter emerge as anti-holomorphic, consisting only of $[..]$ products. Colour-ordered partial amplitudes also have this property. Here however, we are dealing with a physical amplitude, and so the phase is not a measurable quantity. It is interesting to note that choosing the MHV amplitudes to have different phases, for instance an expression containing a mixture of $\lambda$ and $\tilde{\lambda}$, does {\it not} in general lead to correct results for non-MHV amplitudes. This is to be expected, as it is only those amplitudes which, apart from the momentum-conserving delta function, are comprised entirely of $\langle .. \rangle$ products that transform simply onto a line in twistor space \cite{Witten}.

\subsection{The NMHV amplitude $A(\overline{f^+_1}, f^-_2,3^+,4^+,5^-,6^-)$}

The first non-zero NMHV amplitudes appear for $n=4$ photons, when two photons have helicity $+$ and two have helicity $-$.
There are eight diagrams for this process but only three distinct structures, so that we need only work out three diagrams and obtain the others by permuting photons. In fact, by a judicious choice of the arbitrary spinor $\eta$ we can reduce the expression to just two independent terms plus permutations. Referring to Figure~3,
\begin{eqnarray}
A &=& -4\frac{\phi_1^2 [\langle 2 \ 1 \rangle \phi_1 + \langle 2 \ 6 \rangle \phi_6] \langle 2 \ 5 \rangle^2}{[1 \ 6] \phi_6 \langle 2 \ 3 \rangle \langle 2 \ 4 \rangle [\langle 3 \ 1 \rangle \phi_1 + \langle 3 \ 6 \rangle \phi_6][\langle 4 \ 1 \rangle \phi_1 + \langle 4 \ 6 \rangle \phi_6]}\; , \nonumber \\
B &=&  -4\frac{[\langle 6 \ 1 \rangle \phi_1 + \langle 6 \ 4 \rangle \phi_4]^2 \langle 2 \ 5 \rangle ^2}{[\langle 4 \ 1 \rangle \phi_1 + \langle 4 \ 6 \rangle \phi_6][\langle 3 \ 2 \rangle \phi_2 + \langle 3 \ 5 \rangle \phi_5]\langle 1 \ 4 \rangle r^2 \langle 2 \ 3 \rangle} \; ,  \nonumber \\
C &=& -4\frac{[\langle 1 \ 2 \rangle \phi_2 + \langle 1 \ 5 \rangle \phi_5][\langle 6 \ 2 \rangle \phi_2 + \langle 6 \ 5 \rangle \phi_5]^2}{[\langle 3 \ 2 \rangle \phi_2 + \langle 3 \ 5 \rangle \phi_5] [\langle 4 \ 2 \rangle \phi_2 + \langle 4 \ 5 \rangle \phi_5]\langle 1 \ 3 \rangle \langle 1 \ 4 \rangle [2 \ 5] \phi_5}  \; .
\end{eqnarray}
\begin{figure} \label{nmhvdiagrams}
\centering
    \subfigure[Diagram A]{
    \label{fig:subfig:V}
    \includegraphics[width=2.3in]{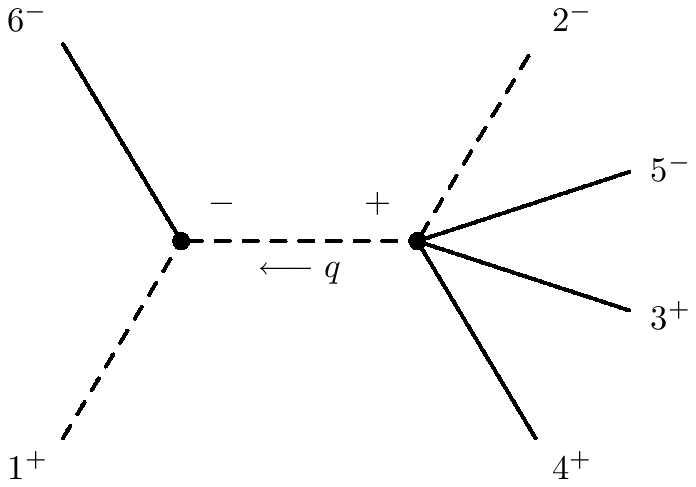}}
    \subfigure[Diagram B]{
    \label{fig:subfig:B}
    \includegraphics[width=2.5in]{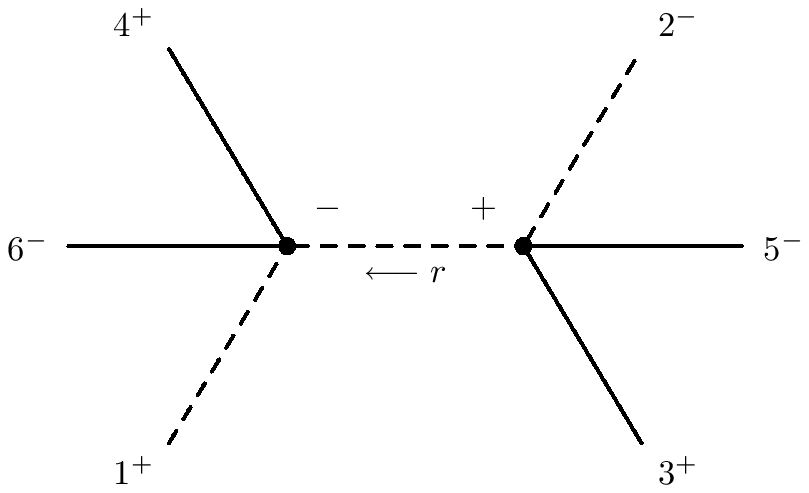}}
    \subfigure[Diagram C]{
    \label{fig:subfig:C}
    \includegraphics[width=2.5in]{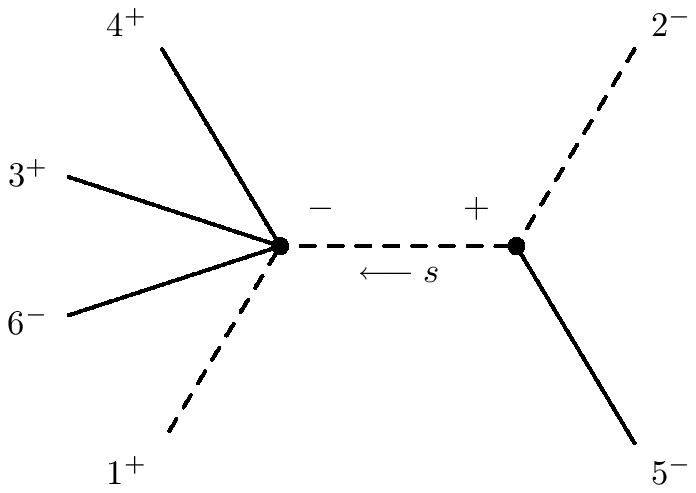}}
\caption{Diagrams contributing to $A(\overline {f
^+_1},f^-_2,3^+,4^+,5^-,6^-)$. Various permutations of each also contribute.}
\end{figure}
If we choose $\eta = \tilde{\lambda_1}$ then $\phi_1 = 0$ and the contribution from diagram A above vanishes. The other two terms simplify, and we end up with the following expression,
\begin{eqnarray} \label{nmhvcsw_result}
  \nonumber A(\overline{f^+_1}, f^-_2,3^+,4^+,5^-,6^-) &=& 4\frac{\langle 4 \ 6 \rangle [ 1 \ 4]^2 \langle 2 \ 5 \rangle^2}{\langle 1 \ 4 \rangle [ 1 \ 6] (p_2 + p_3 + p_5)^2 \langle 2 \ 3 \rangle \langle 3 | 2+5 |1]} \\
 \nonumber  &+& (3 \leftrightarrow 4) + (5 \leftrightarrow 6) +  \left(%
\begin{array}{c}
  3 \leftrightarrow 4 \\
  5 \leftrightarrow 6 \\
\end{array}%
\right)  \\
   \nonumber &+& 4\frac{(s_{12} + s_{15}) \langle 6 | 2+5 | 1]^2}{\langle 1 \ 3 \rangle \langle 1 \ 4 \rangle [ 5 \ 2][1 \ 5] \langle 3|2+5|1] \langle 4|2+5|1]} \\
 &+& ( 5 \leftrightarrow 6) \phantom{\left(%
\begin{array}{c}
  3 \leftrightarrow 4 \\
  5 \leftrightarrow 6 \; .
\end{array}%
\right)}
\end{eqnarray}
The notation $\langle i \ | \ j+k \ | \ l \ ]$ and $s_{ij}$ is defined in Appendix A. Although the two algebraic forms are very different, we have checked that this expression agrees numerically with the KS results \cite{KS}, up to a phase. It is interesting to note that we do not have the freedom to introduce relative phases among the set of MHV amplitudes. For example, introducing a factor of $-1$ into the $1$-photon MHV amplitude while leaving the others fixed will obviously lead to a change in the relative phases among the terms in \eqref{nmhvcsw_result}. Our derived expression for $A(\overline{f^+_1}, f^-_2,3^+,4^+,5^-,6^-)$ will then no longer have the correct magnitude. In this way an apparently unphysical phase affects physical cross sections. So the phases of the MHV amplitudes in \eqref{QEDMHV} must be chosen appropriately.

We have also calculated the NMHV 5-photon amplitude\footnote{Note that all non-zero $n=5$ helicity amplitudes are either MHV or NMHV, as for $n=4$. The first NNMHV amplitudes appear at $n=6$.} $A(\overline{f}^+_1,f^-_2,3^+,4^+,5^-,6^-,7^+)$ using the MHV rules, and once again found numerical agreement with \cite{KS}. We expect that further tests will be successful. Amplitudes with $n$ photons, two of which have negative helicity, require the evaluation of only $n-1$ structures. The full set of diagrams is then easily obtained through permutations. Increasing the number of negative helicity photons leads to MHV diagrams with more than two vertices. For QED processes with two fermions, the absence of a pure-photon vertex means that such diagrams consist only of a linear string of vertices -- there is no branching. Each vertex has one negative helicity photon attached to it, and the remaining photons are added in all possible ways.
\subsection{Soft Limits}

We have checked algebraically that (\ref{nmhvcsw_result}) has the correct limits when one of the photon's momenta is taken to zero, namely that the expression tends to the amplitude in the absence of that photon, multiplied by a `soft' factor called the eikonal factor:
\begin{eqnarray}
A(\overline{f^+_1}, f^-_2,3^+,4^+,5^-,6^-) &\stackrel{3 \rightarrow 0}{\longrightarrow}& A(\overline{f^+_1}, f^-_2,4^+,5^-,6^-) \times \frac{\sqrt{2} \langle 1 \ 2 \rangle}{\langle 1 \ 3 \rangle \langle 2 \ 3  \rangle}  \nonumber \\
A(\overline{f^+_1}, f^-_2,3^+,4^+,5^-,6^-) &\stackrel{5 \rightarrow 0}{\longrightarrow}& A(\overline{f^+_1}, f^-_2,3^+,4^+,6^-) \times \frac{\sqrt{2} [1 \ 2]}{[1 \ 5][ 2 \ 5 ]}
\end{eqnarray}
and similarly for the other photons. Notice that when a positive helicity photon becomes soft, the eikonal factor is comprised entirely of $\langle .. \rangle$ spinor products, whereas when a negative helicity photon becomes soft the eikonal factor is comprised entirely of $[..]$ products.
We have also verified that the MHV amplitudes (\ref{QEDMHV}) have the correct collinear factorization properties when one of the photons is emitted in the direction of the incoming fermion or anti-fermion.\footnote{The collinear behaviour of QCD MHV amplitudes has been studied in \cite{collinear}.}

\section{The BCF Recursion Relations}

A new set of recursion relations \cite{BCF} has been proposed to calculate tree amplitudes in gauge theories. We will here give a brief review of this technique, before showing how the relations can be used, along with (\ref{QEDMHV}), to calculate QED amplitudes.

Consider an $n$ particle (say gluonic, for definiteness) scattering amplitude, with arbitrary helicities. Choose two of the external lines to be `hatted' -- this will be defined shortly. Suppose the $n$-th (positive helicity) and $(n-1)$-th (negative helicity) gluons are hatted. These are reference lines. The BCF recursion relation then reads

\begin{eqnarray} \label{BCF}
& &A_n(1,2,\ldots,(n-1)^-,n^+) =  \nonumber \\
 & &\sum_{i=1}^{n-3} \sum_{h=+,-} A_{i+2}(\widehat{n},1,2,\dots,i,-\widehat{P}^h_{n,i})\frac{1}{P^2_{n,i}} A_{n-i}(+\widehat{P}^{-h}_{n,i},i+1,\dots,n-2,\widehat{n-1})
\end{eqnarray}
where
\begin{eqnarray}
P_{n,i} &=& p_n + p_1 + \dots +p_i \; , \nonumber \\
\widehat{P}_{n,i} &=& P_{n,i} + \frac{P^2_{n,i}}{\langle n-1 | P_{n,i}|n]} \lambda_{n-1}\widetilde{\lambda}_n\; , \nonumber \\
\widehat{p}_{n-1} &=& p_{n-1} - \frac{P^2_{n,i}}{\langle n-1 | P_{n,i}|n]} \lambda_{n-1}\widetilde{\lambda}_n \; , \nonumber\\
\widehat{p}_n &=& p_n + \frac{P^2_{n,i}}{\langle n-1 | P_{n,i}|n]} \lambda_{n-1}\widetilde{\lambda}_n \; .
\end{eqnarray}
Identities such as
\begin{eqnarray}
\langle \bullet \ \widehat{P} \rangle &=& - \langle \bullet \ | \ P \ | \ n] \times  \frac{1}{\omega}\; , \nonumber \\
\lbrack \widehat{P} \ \bullet \rbrack &=& - \langle n-1 \ | \ P \ | \bullet ] \times \frac{1}{\bar{\omega}}\; ,
\end{eqnarray}
where $\omega \overline{\omega} = \langle n-1 \ | P \ | n ]$ (the factors $\omega$ and $\overline{\omega}$ only ever appear together in this combination) are used to remove the hats, whereupon the result can be simplified using standard spinor identities. The procedure can be conveniently represented diagrammatically, see Figure 4 for a specific case.

In Ref.~\cite{LuoWen} the relations were shown to work for amplitudes involving fermions, and in Ref.~\cite{BCFW} it was shown that the reference gluons need not be either adjacent or of the same helicity. Applications to massive particles were described in \cite{massiveBCF}. The relations were proven in \cite{BCFW} by shifting the hatted momenta by a complex amount -- see Appendix B. Here we will be interested in applying the above recursion relation to QED processes. In contrast to the MHV rules, the recursion relations involve the use of $\overline{\mbox{MHV}}$ amplitudes. We can obtain these from (\ref{QEDMHV}) by switching $\langle..\rangle$ and $[..]$, and using charge conjugation invariance to swap the fermion and anti-fermion.

\subsection{Example of BCF recursion relations applied to a QED process}

Consider the MHV amplitude $A(\overline{f}^+_1, f^-_2,3^+,4^+,5^-)$. As before, $f$ and $\overline{f}$ denote a fermion and anti-fermion respectively, and $i^+$ represents a positive helicity photon of momentum $k_i$. Let us choose the hatted lines to be 1 and 5, as shown in Figure 4. Then there is only one distinct diagram\footnote{Note that as detailed in \cite{BCF}, diagrams with an upper vertex of  $(++-)$ or a lower vertex of $(--+)$ vanish. We have not drawn such diagrams.} for this process, which evaluates to
\begin{equation}
\frac{\sqrt{2}[\widehat{1} \ 3]^2}{[\widehat{1} \ \widehat{P}]}  \frac{1}{(k_1 + k_3)^2}  \frac{2\langle 2 \ \widehat{5} \rangle^2}{\langle 2 \ 4 \rangle \langle \widehat{P} \ 4 \rangle} \; .
\end{equation}
\begin{figure}
\centering
\includegraphics[height=2in]{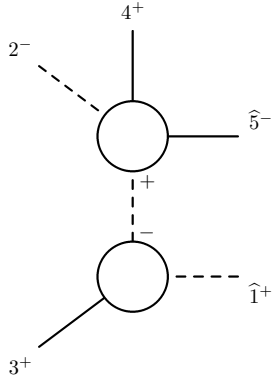}
\caption{BCF Diagram contributing to $A(\overline{f}^+_1, f^-_2,3^+,4^+,5^-)$. As usual, dashed lines are fermions, solid lines are photons.}
\end{figure}
Here we have used (\ref{QEDMHV}), together with its helicity flipped version, to substitute for the (on-shell) tree amplitudes in (\ref{BCF}). $P = k_1 + k_3$ is the momenta of the internal line. Simplifying, we get
\begin{equation}
2\sqrt{2}\frac{\langle 2 \ 5 \rangle^2}{\langle 2 \ 4 \rangle \langle 3 \ 4 \rangle \langle 1 \ 3 \rangle} \; ,
\end{equation}
and to this we must add a similar expression with photons $3$ and $4$ interchanged,
\begin{equation}
2\sqrt{2}\frac{\langle 2 \ 5 \rangle^2}{\langle 2 \ 3 \rangle \langle 4 \ 3 \rangle \langle 1 \ 4 \rangle}\; .
\end{equation}
After simplifying using Schouten's identity\footnote{For any 4 spinors $\langle a \ b \rangle \langle c \ d \rangle + \langle a \ c \rangle \langle d \ b \rangle + \langle a \ d \rangle \langle b \ c \rangle=0$.} we recover the expected result,
\begin{eqnarray}
A(\overline{f}^+_1, f^-_2,3^+,4^+,5^-) &=& 2\sqrt{2}\frac{\langle 2 \ 5 \rangle^2}{\langle 2 \ 4 \rangle \langle 3 \ 4 \rangle \langle 1 \ 3 \rangle}
 + 2\sqrt{2}\frac{\langle 2 \ 5 \rangle^2}{\langle 2 \ 3 \rangle \langle 4 \ 3 \rangle \langle 1 \ 4 \rangle}
\nonumber  \\
&=& 2\sqrt{2}\frac{\langle 2 \ 1 \rangle \langle 2 \ 5 \rangle^3 \langle 1 \ 5 \rangle}{\prod_{k=3}^5 \langle 1 \ k \rangle \langle 2 \ k \rangle}\; .
\end{eqnarray}

\subsection{The NMHV amplitude $A(\overline{f^+_1}, f^-_2,3^+,4^+,5^-,6^-)$}

There are three BCF diagrams for this process, up to permutations, which is the same one that we calculated in Section 4.2 using the MHV rules. We can build up the four photon process using amplitudes we have already calculated. The diagrams (Figure 5) evaluate to
\begin{figure} \label{nmhvbcfdiagrams}
\centering
    \subfigure[Diagram P]{
    \label{fig:subfig:P}
    \includegraphics[height=2in]{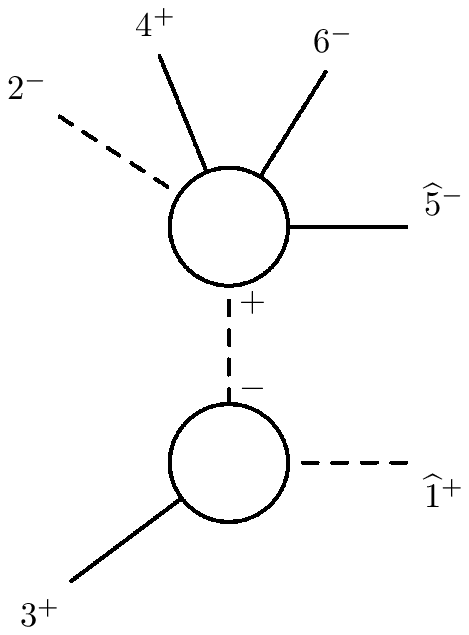}}
    \subfigure[Diagram Q]{
    \label{fig:subfig:Q}
    \includegraphics[height=2in]{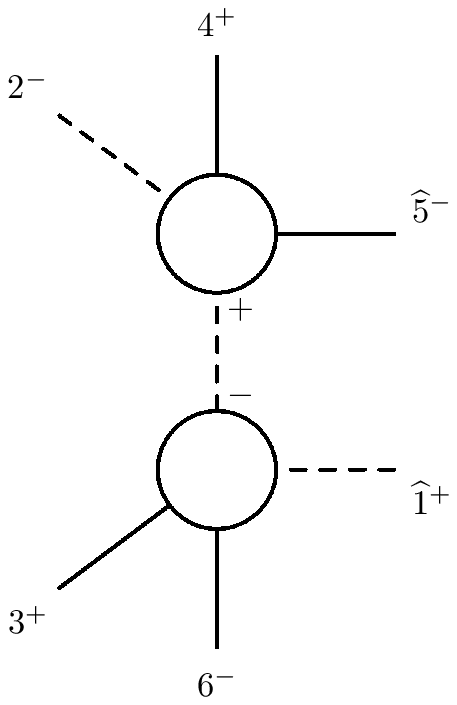}}
    \subfigure[Diagram R]{
    \label{fig:subfig:R}
    \includegraphics[height=1.8in]{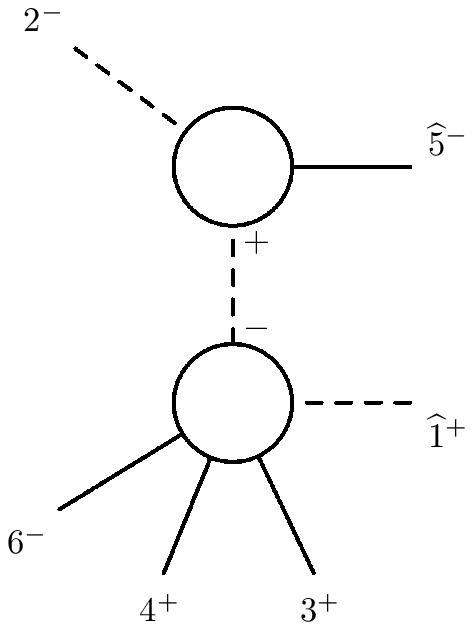}}
\caption{BCF diagrams contributing to $A(\overline {f
^+_1},f^-_2,3^+,4^+,5^-,6^-)$. Various permutations of each also contribute.}
\end{figure}
\begin{eqnarray}
P &=& 4\frac{\langle 5 | 1+3|2] \langle 5|1+3|4]^2}{\langle 3 | 1+5 |2] \langle 5 | 1 +3|6] (p_1 + p_3 + p_5)^2 [2 \ 6] \langle 1 \ 3 \rangle} \; , \nonumber \\
Q &=& 4\frac{\langle 2 \ 5 \rangle ^2 [ 3 \ 1 ]^2 \langle 5|3+6|1]}{\langle 2 \ 4 \rangle \langle 4 | 3+6 | 1] \langle 5|1+3|6] (p_1+p_3+p_6)^2 [6 \ 1]}\; \nonumber  \\
R &=& 4\frac{\langle 6|2+5|1]^2 (p_1+p_2+p_5)^2 [2 \ 1]^2}{\langle 3 | 2 + 5 | 1] \langle 4 | 2 + 5 | 1]\langle 3 | 1 + 5 | 2]\langle 4 | 1 + 5 | 2] [5 \ 1 ] [2 \ 5]} \; ,
\end{eqnarray}
where we have chosen external lines 1 and 5 to be hatted. The full result is then
\begin{equation}
A(\overline{f^+_1}, f^-_2,3^+,4^+,5^-,6^-) = (P + Q) + (3 \leftrightarrow 4) + R\; .
\end{equation}
We have checked numerically that this expression is equal to (\ref{nmhvcsw_result}) which was calculated using the MHV rules. Both are equal to the corresponding result obtained from the KS formula, up to a phase.

\section{Conclusions}
We have shown that the modern techniques inspired by the transformation of Yang-Mills scattering amplitudes to twistor space \cite{Witten} can be successfully applied to QED processes, and yield reasonably compact expressions. As well as some simple $\overline{\mbox{MHV}}$ amplitudes, we calculated the NMHV amplitude $A(\overline{f^+_1}, f^-_2,3^+,4^+,5^-,6^-)$ using both the MHV rules and BCF recursion approaches. The expressions obtained are not obviously equal, but numerical checks proved them to be so and the results were confirmed by comparison with the KS \cite{KS} formula, which is directly derived from Feynman diagrams. We have also checked that the
amplitudes have the correct factorising (eikonal) form when one of the photons becomes soft. Note that the QED NMHV amplitudes we have presented can also in principle be obtained by symmetrizing QCD colour-ordered amplitudes, but this is a laborious procedure and will not lead directly to compact expressions. We have shown that it is possible, and much easier, to use {\it physical} MHV amplitudes directly in the MHV rules.

We have given explicit expressions for up to and including 4-photon amplitudes. The extension to $ n \ge 5$ photons is in principle straightforward -- in either the CSW or BCF approaches -- although there is an inevitable growth in complexity as more
N$^n$MHV amplitudes start to appear. We have not been able to discern any large-$n$ simplification of the expressions, in contrast to the remarkably compact expression for arbitrary $n$ (see Eq.~(\ref{eq:ks1})) in the KS approach.

One technical point deserves comment. It turns out that defining the phases of the MHV amplitudes is not a trivial matter. As may have been expected, it is necessary to choose them to be holomorphic functions of the $\langle..\rangle$ spinor product. Even then, unphysical relative phases among the set of MHV amplitudes influence observable results such as the absolute values of derived non-MHV amplitudes. The choices made in (\ref{QEDMHV}) work for all the amplitudes we have calculated, but we are unable to motivate them in a convincing way. One way is to define the one photon vertex with an arbitrary phase (though still holomorphic) and then use the BCF recursion relations to derive all other vertices. Our choice conforms to this.

Finally, it should be obvious that {\it any} QED amplitude can be built up in a similar way. Of particular practical interest, for example, are the amplitudes for processes with four fermions, $e^+e^- \to \mu^+\mu^- + n \gamma$.  Results for these will be presented in a future publication \cite{WIP}.

\vskip 1cm
\noindent{\bf Acknowledgements} \\\\
KJO acknowledges the award of a PPARC studentship. We are grateful to Valya Khoze for useful discussions.

\newpage
\section*{Appendices}
\appendix
\section{Notation and Conventions}

We use the spinor helicity formalism \cite{Spinor}, in which on-shell momenta of massless particles are represented as
\begin{equation}
p_{a\dot{a}} = \lambda_a \widetilde{\lambda}_{\dot{a}} \; .
\end{equation}
Here $\lambda_a$ and $\tilde{\lambda}_{\dot{a}}$ are commuting spinors with positive and negative chirality respectively. We define two types of spinor product:
\begin{equation}
\langle \lambda \ \lambda' \rangle = \epsilon_{ab} \lambda^a \lambda ^b
\end{equation}
and
\begin{equation}
[ \tilde{\lambda} \ \tilde{\lambda}' ] = \epsilon_{\dot{a}\dot{b}} \tilde\lambda^{\dot{a}} \tilde\lambda^{\dot{b}} \; .
\end{equation}
If we have two null 4-vectors $p_{a\dot{a}} = \lambda_a \tilde{\lambda}_{\dot{a}}$ and $q_{a\dot{a}} = \lambda_{a}' \tilde{\lambda'}_{\dot{a}}$ then
\begin{equation} \label{spinordef}
s_{pq}= 2p \cdot q = \langle \lambda \ \lambda' \rangle [\tilde{\lambda} \ \tilde{\lambda}'] \; .
\end{equation}
For clarity, we will usually abbreviate the notation by writing the spinor products as
\begin{eqnarray}
\langle \lambda_i \ \lambda_j \rangle &=& \langle i \ j \rangle\; , \\
\lbrack \tilde{\lambda}_i \ \tilde{\lambda}_j ] &=& [ i \ j ]\; .
\end{eqnarray}
Also, it is useful to define $\langle i | j+k | l ] = \langle i \ j \rangle [j \ l] + \langle i \ k \rangle [ k \ l]$.
For amplitudes considered here we take all particles to be incoming so that, in an $n$ particle process, $p_1 + p_2 +\dots+ p_n = 0$ and
\begin{equation}
\sum_i^n \langle j \ i \rangle [ i \ k ] = 0
\end{equation}
for all $j,k$.

In practical applications the spinor product $\langle i \ j \rangle$  can be conveniently represented in terms of the 4-momenta components \cite{ManganoParke},
\begin{equation}
\langle i \ j \rangle = \frac{(k_i^1 k_j^+ - k_j^1 k_i^+)}{\sqrt{k_i^+ k_j^+}} +i\frac{(k_i^2 k_j^+ - k_j^2 k_i^+)}{\sqrt{k_i^+ k_j^+}}\; ,
\end{equation}
where $k^{\pm} = k^0 \pm k^3$. The square bracket $[i \ j]$ product is then defined using Eq.~(\ref{spinordef}). For a thorough review of the spinor helicity formalism, the reader is directed to Refs.~\cite{ManganoParke,Dixon}.

\newpage

\section{Proof of the Recursion Relations}

An elegant proof of the relations proposed in \cite{BCF} was presented in \cite{BCFW}. Here we will briefly sketch its main elements, before discussing its applicability to QED.

Take a tree level amplitude $A(1,2,\dots,n)$ with arbitrary helicities and
\begin{itemize}
\item choose two particles for special treatment, which we can take to be the $k$-th and $l$-th particles with helicities $h_k$ and $h_l$ respectively, and introduce a complex variable $z$ to rewrite their momenta as
\begin{eqnarray}
p_k(z) &=& \lambda_k (\tilde{\lambda}_k - z\tilde{\lambda}_l) = p_k(0) - z\lambda_k\tilde{\lambda}_l \; ,\nonumber \\
p_l(z) &=& (\lambda_l +z\lambda_k)\tilde{\lambda}_l = p_l(0) + z\lambda_k\tilde{\lambda}_l \; .
\end{eqnarray}
We have effectively shifted the spinors $\lambda_l \rightarrow \lambda_l + z\lambda_k$ and $\tilde{\lambda}_k \rightarrow \tilde{\lambda}_k - z\tilde{\lambda}_l$. Note that there is no symmetry between $k$ and $l$ -- they are treated differently. Having done this we can now construct the auxiliary function
\begin{equation}
A(z) = A(p_1,p_2,\dots,p_k(z),\dots,p_l(z),\ldots,p_n) \; .
\end{equation}
The aim now is to use the analytic structure of this auxiliary function, considered as a function of $z$.
\item $A(z)$ has only simple poles. This can be argued by noting that poles only arise from propagators $1/K^2$, where $K$ is the momenta of the internal line. If both $p_l$ and $p_k$, or neither of them, are present in the sum of external momenta contributing to $K$ then the latter is independent of $z$ and there is no $z$-pole in the propagator. However, if only one and not the other is present then the momenta of the internal line is linearly dependent on $z$, and so is the propagator. Thus $A(z)$ has only simple poles.
\item
Cauchy's theorem tells us that
\begin{equation}
A(0) = -\sum_{\alpha} \textrm{Residue} \left( \frac{A(z)}{z} \right)_{z=z_{\alpha}}\ - \ \textrm{Residue} \left( \frac{A(z)}{z}\right)_{z=\infty}
\end{equation}
so that the physical amplitude $A(0)$ is fully determined by the finite pole positions $z_\alpha$ and residues of the auxiliary function, provided $A(z)$ vanishes at infinity. The finite residues are just products of lower-$n$ tree amplitudes, with Feynman propagators in between. The recursion relation then follows immediately.
\end{itemize}

To demonstrate the vanishing of $A(z)$ as $z \rightarrow \infty$, one may use the MHV rules outlined in \cite{CSW}. It suffices to show that the MHV amplitudes themselves vanish in this limit since, as shown in \cite{BCFW}, the off-shell continuation does not affect the large $z$ behaviour of a general MHV diagram. It turns out that some choices of reference lines are allowed (i.e. lead to an auxiliary function that vanishes at infinity), whilst others are not. We can formulate some rules to determine the allowed choices. This is useful because, as the authors of Ref.~\cite{LuoWen} found, the number of BCF diagrams contributing to a given amplitude depends strongly on the reference lines chosen. A careful choice can save much labour, and yield more compact expressions.

First, let us repeat (\ref{QEDMHV}) for convenience,
\begin{equation} \label{QEDMHV2}
A(\overline{f}^+,f^-,1^+,2^+...,I^-,...,n^+) =
\frac{2^{\frac{n}{2}} \langle f \overline{f} \rangle^{n-2} \langle
f I \rangle^3 \langle \overline{f} I \rangle}{\prod_{i=1}^n
\langle f i \rangle \langle \overline{f} i \rangle}\; ,
\end{equation} and also its $\overline{\mbox{MHV}}$ counterpart
\begin{equation} \label{MHVbar}
A(\overline{f}^+,f^-,1^-,2^-,...,I^+,...,n^-) \doteq
\frac{2^{\frac{n}{2}} [ \overline{f} f ]^{n-2} [
\overline{f} I ]^3 [ f I]}{\prod_{i=1}^n
[ f i ] [ \overline{f} i]}\; .
\end{equation}
The MHV rules can be employed using either solely MHV or solely $\overline{\mbox{MHV}}$ amplitudes. If we choose $l$ to be a positive helicity photon, and consider (\ref{QEDMHV2}) then it is clear that the amplitude vanishes at infinity since there are more factors of $z$ in the denominator than in the numerator. This is true regardless\footnote{In fact if $k$ is either the fermion or antifermion, then $A(z)$ vanishes as $1/z$ whereas if $k$ is another photon then $A(z)$ vanishes as $1/z^2$.} of the identity of $k$. Similarly if we choose $k$ to be a negative helicity photon, and consider (\ref{MHVbar}) then once again $A(z)$ vanishes at infinity, regardless of our choice of $l$. So in both these cases, which cover a large subset of the possible choices, the recursion relations will work. The positive helicity anti-fermion may be used at the lower vertex provided the fermion is not used at the upper vertex, as in this case the MHV amplitude does not vanish as $z \to \infty$.

It is also possible \cite{BCFW} to see the analytic structure of an amplitude by considering the set of Feynman diagrams that contribute to it. For $e^+e^- \to n\gamma$ there are $n!$ diagrams, differing only in the order in which the photons are attached to the fermion line. The $z$-dependence of the diagram can only come from propagators (which either contribute a factor $1/z$ or are independent of $z$) and photon polarization vectors\footnote{In contrast to QCD, the vertices are momentum independent and so cannot depend on $z$.} which, in the spinor helicity formalism, take the general form
\be
\epsilon^-_{a\dot{a}} = \frac{\lambda_a \tilde{\mu}_{\dot{a}}}{[\tilde{\lambda} \ \tilde{\mu}]}, \; \; \; \;
\epsilon^+_{a\dot{a}} = \frac{\mu_a \tilde{\lambda}_{\dot{a}}}{\langle \mu \ \lambda \rangle}
\ee
for negative and positive helicity photons respectively. Here $\mu$ and $\tilde{\mu}$ are reference spinors. Recall that we shift the spinors representing the momenta of the $l$-th and $k$-th legs as
\bea
\lambda_l &\rightarrow& \lambda_l + z\lambda_k \nonumber \\
\tilde{\lambda}_k &\rightarrow& \tilde{\lambda}_k - z\tilde{\lambda}_l
\eea
so that the polarization vector of the $k$-th photon behaves as $1/z$ if it has negative helicity and linearly in $z$ if it has positive helicity. The opposite holds for the $l$-th photon. By looking at the most dangerous Feynman graphs we can deduce that choosing $h_k=-$ or $h_l=+$ is always allowed, which verifies what we saw above using MHV diagrams.

\end{document}